\documentclass[aps,prl,twocolumn,groupedaddress,showpacs]{revtex4}
\usepackage{natbib}
\usepackage{graphicx}
\usepackage{epsfig}
\usepackage{amsmath}

\begin{document}

\title{Dark Compact Planets}

\author{Laura Tolos} 
\affiliation{Instituto de Ciencias del Espacio
  (IEEC/CSIC), Campus Universitat Autonoma de Barcelona, Carrer de Can
  Magrans, s/n, 08193 Cerdanyola del Valles, Spain; \\
  Frankfurt Institute for Advanced Studies. J. W. Goethe Universit\"at,
  Ruth-Moufang-Str.~1, 60438 Frankfurt am Main, Germany}

\author{J\"urgen Schaffner-Bielich} 
\affiliation{Institut f\"ur Theoretische Physik, J. W. Goethe Universit\"at, 
  Max von Laue-Str.~1, 60438 Frankfurt am Main, Germany}

\date{\today}


\begin{abstract}
  We investigate compact objects formed by dark matter admixed with ordinary
  matter made of neutron star matter and white dwarf material. We consider
  non-self annihilating dark matter with an equation-of-state given by an
  interacting Fermi gas. We find new stable solutions, dark compact planets, with Earth-like masses and radii from few Km to few hundred Km for weakly interacting dark matter which are stabilized by the mutual presence of dark matter and compact star matter. For the strongly
  interacting dark matter case, we obtain dark compact planets with
  Jupiter-like masses and radii of few hundred Km.  These objects
  could be detected by observing exoplanets with unusually small radii.
  Moreover, we find that the recently observed 2 ${\rm M}_{\odot}$ pulsars set
  limits on the amount of dark matter inside neutron stars which is, at most,
  $10^{-6}{\rm M}_\odot$.
\end{abstract}

\maketitle

Astrophysical and cosmological observations reveal that most of the mass of
the universe is in the form of dark matter (DM)
\cite{Ade:2013zuv,Betoule:2014frx}. The nature of dark matter is still
elusive. There are direct methods for detecting DM using particle accelerators
\cite{ATLAS:2012ky,Chatrchyan:2012me} or analyzing DM scattering off nuclear
targets in terrestrial detectors \cite{Klasen:2015uma}, such as CDMS-II \cite{Ahmed:2010wy},
CREST-II \cite{Angloher:2011uu}, CoGENT \cite{Aalseth}, DAMA/LiBRA
\cite{Bernabei:2010mq}, LUX \cite{Akerib:2013tjd}, SIMPLE
\cite{Felizardo:2011uw}, XENON10/100 \cite{Essig-Aprile}, and
SuperCDMS \cite{Agnese:2013jaa}.

Apart from these direct searches, constraints on the properties of DM can be
extracted by studying DM stars \cite{Dai:2009ik,Kouvaris:2015rea} or the effects of DM on compact objects, such as white
dwarfs  and neutron stars. Neutron stars are the densest
observable objects in the universe and, thus, a privileged laboratory for the
exploration of matter under extreme conditions. The structure of neutron stars
is determined by the equation-of-state (EoS), which is well constrained up to
normal nuclear saturation density \cite{Tews:2012fj}.  However, the properties
of matter at supranormal densities are unknown while being fundamental for the
determination of the maximum mass of neutron stars. Recent accurate
measurements of $2{\rm M}_{\odot}$ neutron stars 
\cite{Demorest:2010bx,Antoniadis:2013pzd} are setting some tension among
different determinations of EoS and, hence, the structure of neutron stars.

The possible gravitational collapse of a neutron star due to accretion of DM can set
bounds on the mass of DM candidates
\cite{Goldman:1989nd,Kouvaris:2011fi}. Also, constraints on DM can be obtained
from stars that accrete asymmetric DM during their lifetime and then collapse
into a white dwarf or neutron star, inheriting the accumulated DM
\cite{Kouvaris:2010jy}. Moreover, the cooling process of compact objects can
be affected by the capture of DM, which subsequently annihilates heating the
star
\cite{Kouvaris:2007ay,Bertone:2007ae,Kouvaris:2010vv,McCullough:2010ai,deLavallaz:2010wp}. At
the same time, self-annihilating DM accreted onto neutron stars may affect significantly
their kinematical properties \cite{PerezGarcia:2011hh} or provide a mechanism
to see compact objects with long-lived lumps of strangelets
\cite{PerezGarcia:2010ap}. Therefore, it is of high interest to analyze the
effects of DM on compact stellar objects.

Recently, neutron stars with non-self annihilating DM have emerged as an interesting
astrophysical scenario, where to analyze the gravitational effects of DM onto
ordinary matter (OM) under extreme conditions
\cite{deLavallaz:2010wp,Li:2012ii,Sandin:2008db,Leung:2011zz,Xiang:2013xwa,Goldman:2013qla}.
In \cite{Sandin:2008db} the two-fluid system of mirror DM
and normal matter coupled through gravity was considered. An admixture of
degenerate DM with masses of $\sim$1 GeV and normal matter was studied in
\cite{Leung:2011zz} using a general relativistic two-fluid formalism. Also, in
\cite{Xiang:2013xwa} the approach of \cite{Sandin:2008db}
was followed with various masses in the GeV range for DM and different interactions. DM
admixed white dwarfs were discussed in \cite{Leung:2013pra} assuming an ideal degenerate
Fermi gas for non-self annihilating DM.

In this paper we investigate compact objects formed by DM admixed with neutron star
matter but also with white dwarf material. We solve the two-fluid system of DM and
OM coupled gravitationally while performing, for the first time, a stability analysis of the DM admixed
objects for non-self annihilating DM with a particle mass of 100 GeV \cite{Spergel:1999mh,Harvey:2015hha}.
 We find new stable solutions of the Tolman-Oppenheimer-Volkoff (TOV)
equations with planet-like objects of Earth-like masses and radii from few Km
to few hundred Km for weakly interacting DM. For the strongly interacting DM
case, we obtain planets with Jupiter-like masses and radii of few hundred Km.
These dark compact planets could be detected via exoplanet searches \cite{Borucki:2010zz} or
using gravitational microlensing \cite{Alcock:1995zx,Sumi:2011kj,
  Udalski:2004dy,Cassan:2012jx}. The detection of an unsually small radius
implying an enormously high mass density would reveal a compact planet, much
in the same way as the first detection of white dwarfs. Finally, we explore
the DM content that neutron stars of 2 ${\rm M}_{\odot}$
\cite{Demorest:2010bx,Antoniadis:2013pzd} might sustain.

In order to analyze the dark compact objects formed by the admixture of
OM with DM, one has to solve the TOV equations for
a system of two different species that interact gravitationally. One proceeds
by solving the dimensionless coupled TOV equations \cite{Narain:2006kx} for OM
and DM,
\begin{eqnarray}
&&\frac{dp'_{OM}}{dr}=-(p'_{OM}+\rho'_{OM})\frac{d \nu}{dr}, \nonumber \\
&&\frac{dm_{OM}}{dr}=4 \pi r^2 \rho'_{OM}, \nonumber \\
&&\frac{dp'_{DM}}{dr}=-(p'_{DM}+\rho'_{DM}) \frac{d \nu}{dr}, \nonumber \\
&&\frac{dm_{DM}}{dr}=4 \pi r^2 \rho'_{DM}, \nonumber \\
&&\frac{d \nu}{dr}=\frac{(m_{OM}+m_{DM}) + 4 \pi r^3(p'_{OM}+p'_{DM})}{r(r-2(m_{OM}+m_{DM}))}, 
\end{eqnarray}
where $p'$ and $\rho'$ are the dimensionless pressure and energy density,
respectively, defined as $p'=P/m_f^4$ and $\rho'=\rho/m_f^4$, with $m_f$ being
the mass of the fermion, that is the neutron mass and dark matter particle
mass, respectively.  The physical mass and radius for each species are $R=
 (M_p/m_f^2) \, r$ and $M= (M_p^3/m_f^2) \,  m$, respectively, where $M_p$ is the Planck mass \cite{Narain:2006kx}.

For OM, we consider neutron star matter given by the equation-of-state EoSI
from \cite{Kurkela:2014vha}. The EoSs obtained in \cite{Kurkela:2014vha} are
constrained by using input from low-energy nuclear physics, the high-density
limit from perturbative QCD, and observational pulsar data. In particular, the EoSI
is the most compact one with maximum masses of 2${\rm M}_{\odot}$. We, moreover,
map EoSI with an inner and outer crust EoS using \cite{Negele:1971vb} and
\cite{Ruester:2005fm}, respectively. For $\rho < 3.3 \times 10^{3}$
$\rm{g/cm^3}$ we use the Harrison-Wheeler EoS \cite{harrison-wheeler}. For DM,
the EoS is taken from \cite{Narain:2006kx} for a non-annihilating massive
DM particle of 100 GeV,  in line with several scenarios for DM, such as asymmetric
DM, that involve massive particles which cannot annihilate with themselves
\cite{Nussinov:1985xr,Kaplan:1991ah,Hooper:2004dc,Kribs:2009fy,Kouvaris:2015rea}.
Two cases are studied, weakly and strongly interacting DM. The strength of the interaction is measured by the strength parameter, $y=m_f/m_I$, which is defined as the ratio between the fermion mass $m_f$ and the interaction mass scale $m_I$ (see Eqs.~(34-35) in Ref.~\cite{Narain:2006kx}). For strong interactions, $m_I \sim 100~{\rm MeV}$  (interaction scale related to the exchange of vector mesons) while for weak interactions $m_I \sim 300~{\rm GeV}$ (exchange of W and Z bosons). Thus, we take $y=10^{-1}$ and $y=10^{3}$ for weakly and strongly interacting DM, respectively.

\begin{figure}[!t]
\includegraphics[width=6cm,height=8.5cm]{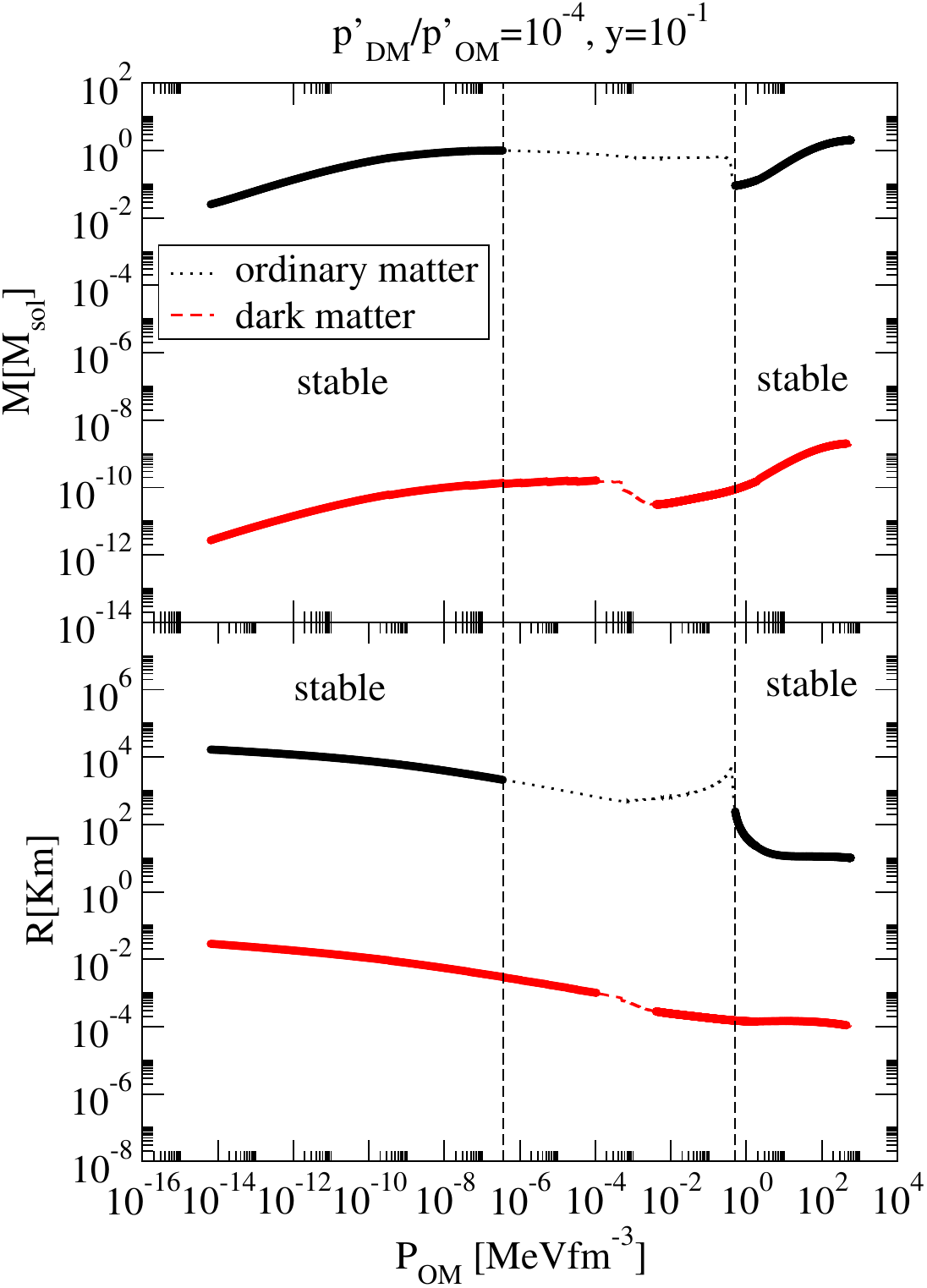}
\caption{Mass and radius of OM and DM objects as a function of the pressure of OM. The solid lines indicate
  the stable  regions for OM and DM, separately. The vertical
  dashed lines delimit the common stable regions for both species. The
  calculation is performed for the weakly interacting DM case, with a strength
  parameter $y=10^{-1}$, and the ratio of pressures between DM and OM is equal
  to $10^{-4}$.}
\label{stability}
\end{figure}

For the determination of the dark  compact objects, one first has to perform an
analysis of the stable configuration for both OM and DM.  The stability arguments can be found, for example, in \cite{Shapiro:1983du}, where the
stability of the different radial modes in a star is analyzed. However, we present here a short description of the stability criteria.
We consider small radial perturbations of the equilibrium configuration, which leads to a Sturm-Liouville eigenvalue equation  \cite{Shapiro:1983du}.
By solving this equation, we find that the eigenfrequencies of the different modes form a discrete hierarchy $\omega_n^2 < \omega_{n+1}^2$ with n=0,1,2..., being $\omega_n^2$  real numbers. A negative value of $\omega_n^2$ leads to an exponential growth of the radial perturbation and collapse of the star. The determination of the sign of the mode results from the analysis of the mass of the star versus the mass density or radius \cite{Shapiro:1983du}. The extrema in the mass versus mass density (pressure) indicates a change of sign of the eigenfrequency associated to a certain mode and, hence, a change of  stability of the star for a given mass density (pressure). The derivative of the radius versus the mass density at that mass
density  determines if the change of stability takes place for an even mode (negative derivate) or for an odd mode (positive
derivative).  Thus, starting at low mass densities where all modes are positive, one can perform the stability analysis for higher mass densities studying the change of sign of the different modes while keeping the hierarchy among them. Only when all eigenfrequencies are positive, the star will be stable. In this way, one can study the stable regions
for both OM and DM. 

In Fig.~\ref{stability} we present this analysis for a
ratio of the dimensionless central pressures for DM and OM equal to
$p'_{DM}/p'_{OM}=10^{-4}$ and the weakly interacting DM case ($y=10^{-1}$) as
a function of the OM central pressure. We show with solid lines the stable
mass-radius regions for OM and DM, separately, while the vertical dashed lines
delimit the common stable regions for both species (denoted with "stable" legend). We distinguish two common
stable mass-radius regions, one for OM central pressures below $10^{-6} {\rm
  MeV fm}^{-3}$ and another one for OM central pressures above $10^{-1} {\rm
  MeV fm}^{-3}$. A similar analysis has been done for other 
ratios of pressures and the strongly interacting DM case, obtaining similar results.

\begin{figure}[!t]
\includegraphics[width=0.5\textwidth,height=7cm]{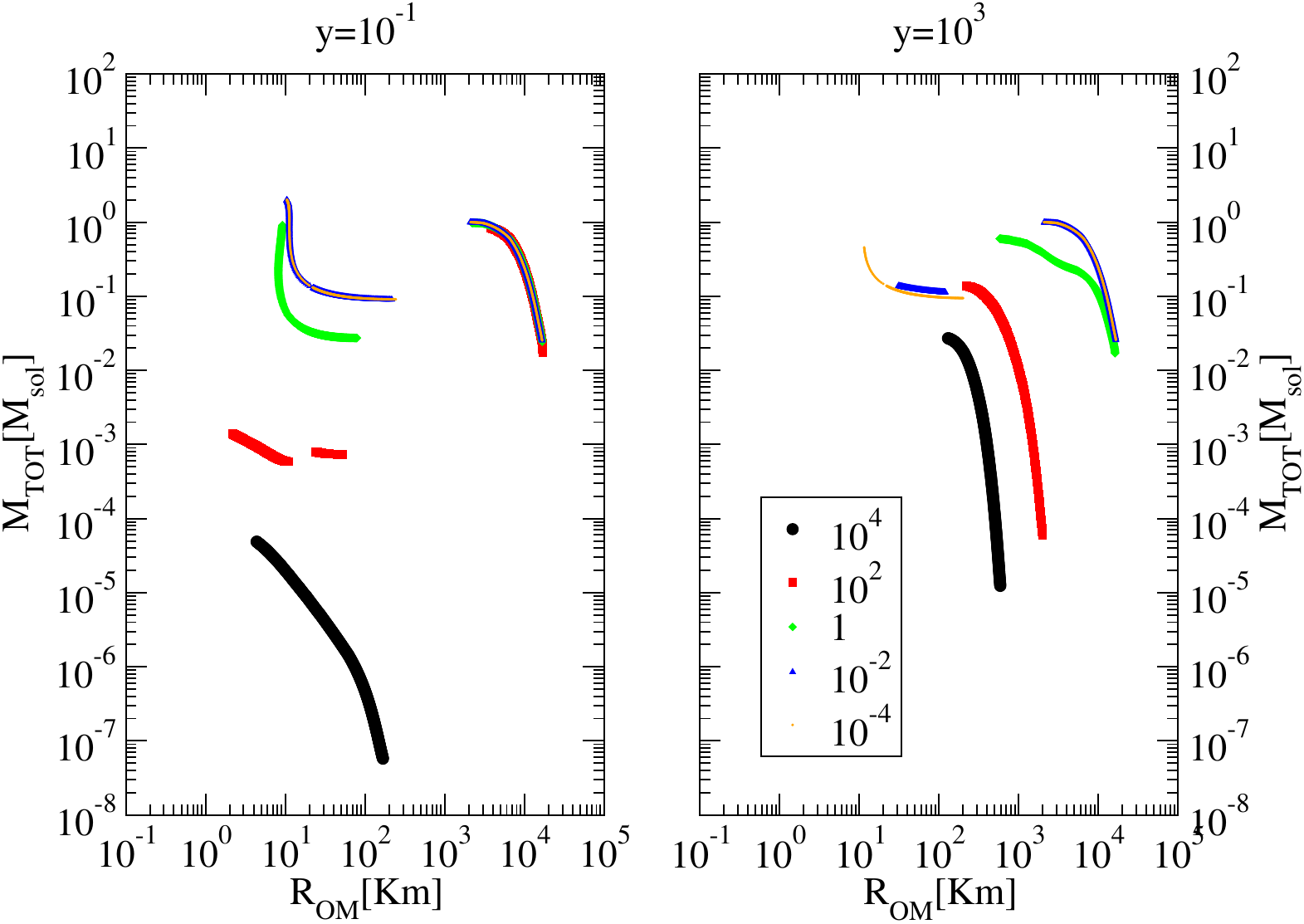}
\caption{The total mass of the dark compact object as a function of the
  observable radius (radius of OM) for different ratios of 
  pressures between DM and OM. Two different strengths
  for the interacting DM are used: $y=10^{-1}$ (left column) and $y=10^3$
  (right column).}
\label{rnm_mtot}
\end{figure}

Once the stability analysis is performed, we study the structure of the
obtained dark compact objects, calculating the total mass of the object as a
function of the visible radius, that is, the radius of OM, in an equivalent
manner as done for neutron stars or white dwarfs. Neutron stars and white dwarfs exhibit
different mass-radius relationships due to their composition and, hence,
EoS. While neutron stars have typical masses of 1-2 ${\rm M}_{\odot}$ and radius of 10 Km,
white dwarfs are characterised by masses of 1${\rm M}_{\odot}$ and radii of few
thousand Km.

In Fig.~\ref{rnm_mtot} we present the mass-radius relationships of the dark
compact objects for different values of the ratio of DM versus OM dimensionless
central pressures as well as the weakly and strongly interacting DM cases. For
a small ratio of $p'_{DM}/p'_{OM}=10^{-4}$, we observe two stable mass-radius
configurations, that correspond to the neutron-star and white-dwarf branches
of OM. As the central pressure of DM increases, the neutron-star branch
becomes unstable and we are left with the white-dwarf branch, with densities
for OM below the neutron drip line but unconventional masses and radii. Indeed, for
the weakly interacting DM case and $p'_{DM}/p'_{OM}=10^{4}$, we obtain a new type
of objects, hereafter named dark compact planets (DCPs), with Earth-like masses
of $M=$$10^{-4} -10^{-7}$${\rm M}_{\odot}$ and radii from few Km to few hundred
Km. As for the strongly interacting case, these DCPs can reach masses similar
to the Jupiter mass, ranging from $10^{-2}-10^{-5}$${\rm M}_{\odot}$, and have radii
of few hundred Km.  The DM content of these DCPs ranges from $M=$$10^{-3}
-10^{-8}$${\rm M}_{\odot}$ for Earth-like planets, and $M=$$10^{-1}
-10^{-5}$${\rm M}_{\odot}$ for Jupiter-like ones.  Interestingly, by choosing a mass of few GeVs for the DM particle, we obtain DCPs of similar size but much more massive, thus, more compact.  Note that by rescaling the EoS for the crust with the iron mass, we reproduce similar results for our DCPs if we increase the pressure of the dark matter relatively to the pressure of ordinary matter.  Also, by
taking the iron mass instead of neutron mass, it will give a shift in the maximum mass of white dwarfs of the order of the binding energy per baryon relative to the nucleon mass, i.e. below $1\%.$ 

At this point, one should consider the different possibilities of having
compact stellar objects with DM content, that is, by means of the accretion
mechanisms of DM onto neutron stars and white dwarfs as well as by the primordial formation of DM
clumps surrounded by OM due to free streaming. On one hand, one
can estimate the DM accreted in neutron stars and white dwarfs by using the accretion rate of heavy
DM particles onto a neutron star \cite{PerezGarcia:2010ap,Kouvaris:2010jy} and a white dwarf
\cite{Kouvaris:2010jy}. A maximum limit for the total accreted mass of DM of
$10^{-14}$${\rm M}_{\odot}$ is obtained for neutron stars, whereas in white dwarfs the maximum total DM mass
accreted is $10^{-12}$${\rm M}_{\odot}$. Thus, the accretion mechanism cannot
explain the existence of a DCP.  Free streaming estimates of 1 pc
\cite{Gorbunov:2011zzc}, on the other hand, give a lower limit for a total DM
mass of $10^{-7}$${\rm M}_{\odot}$. Interestingly, limits on primordial DM objects
from Kepler \cite{Griest:2013aaa} and by the Experience pour la Recherche d'Objets Sombres (EROS) collaboration
\cite{Tisserand:2006zx} rule out the mass range of $2\times10^{-9}$ to
$10^{-7}$${\rm M}_{\odot}$ and $0.6\times10^{-7}$ to $15{\rm M}_\odot$, respectively, to constitute the
entirety of DM in the Milky Way. However, it is still not ruled out that a
fraction of DM form primordially dark compact objects and accrete white dwarf
material subsequently, thereby producing a DCP that becomes
observable with visible light. A detailed analysis of the primordial formation of our DCPs will be done elsewhere \cite{future}.

The search of a DCP could be done in a similar manner as for extrasolar
planets (exoplanets). Most of them have been discovered using radial velocity
or transit methods.  The observation of a DCP using these techniques is provided
by measuring an unusually small upper limit to the radius from transiting
planets. Gravitational microlensing, as
seen by the MAssive Compact Halo Object (MACHO) project
\cite{Alcock:1995zx}, the Microlensing Observations in
Astrophysics (MOA) \cite{Sumi:2011kj}, the Optical Gravitational Lensing
Experiment (OGLE) \cite{Udalski:2004dy} or EROS \cite{Tisserand:2006zx}, could
be an alternative tool for detecting a DCP.

\begin{figure}[!t]
\includegraphics[width=6cm,height=6cm]{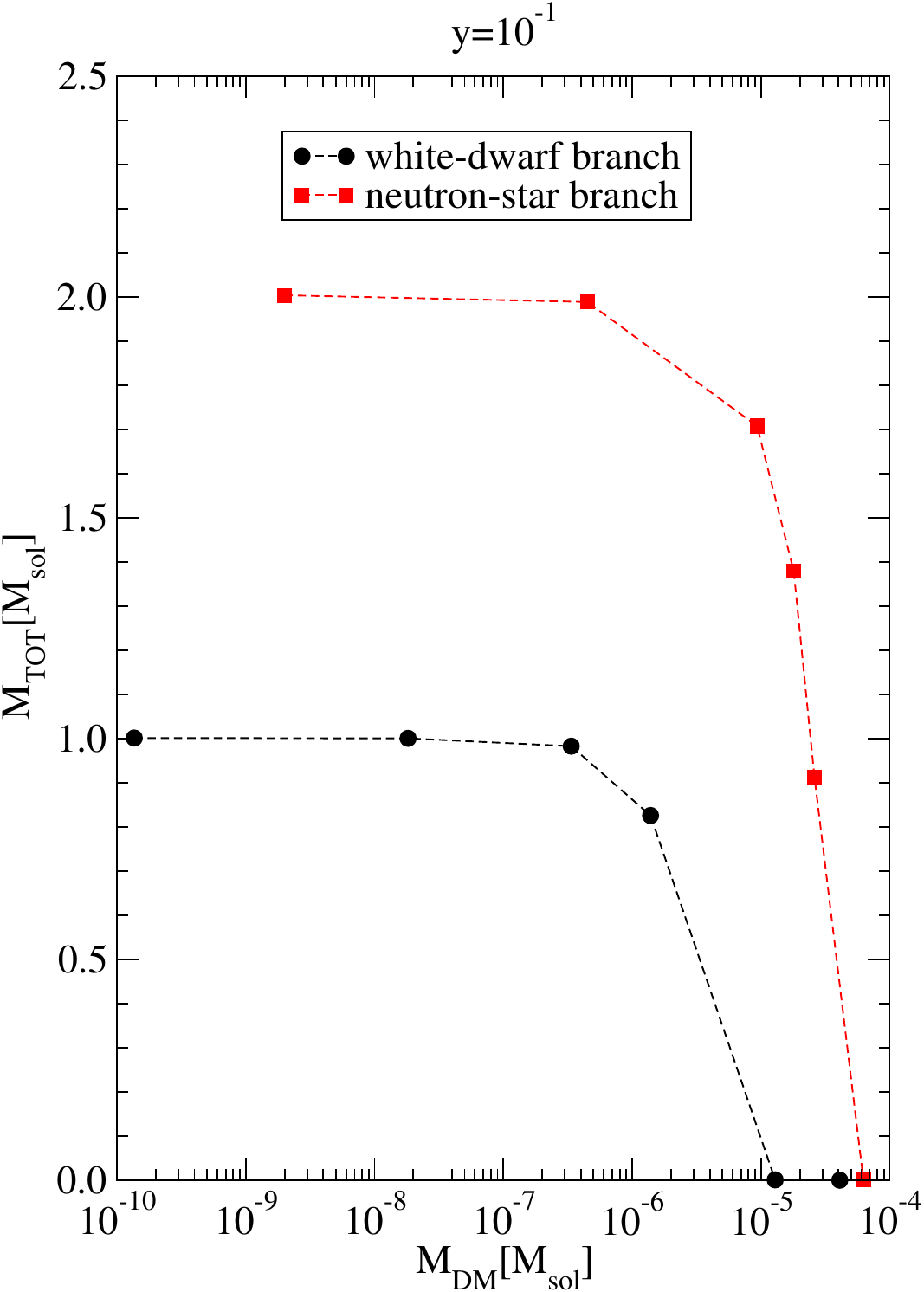}
\caption{The total mass of the dark compact object as a function of the mass
  of DM for weakly interacting DM, $y=10^{-1}$. The full squares correspond to the neutron-star branch, while the full circles indicate the white-dwarf branch. The
  dotted lines match the different symbols given by the actual calculations
  for different ratios of pressures between DM and OM.}
\label{mtot_mdar}
\end{figure}

It is also interesting to study the effect of DM on the 2${\rm M}_{\odot}$ recent
observations of neutron stars \cite{Demorest:2010bx,Antoniadis:2013pzd}, in order to
determine the amount of DM that such massive neutron stars can sustain.  In
Fig.~\ref{mtot_mdar}, we display the total mass of the dark compact object as a function of
the mass of DM for the strength parameter $y=10^{-1}$ and for the
neutron-star and white-dwarf branches. We observe that as the DM
content reaches masses beyond $10^{-6}$${\rm M}_{\odot}$, a neutron star with 2 ${\rm M}_{\odot}$
can not be obtained. There is even a more stringent limit in the DM content
coming from the reduction of the nominal mass of white dwarf of 1${\rm M}_{\odot}$.  Similar
results for the white-dwarf branch are found for the strongly interacting
DM case, while the maximum mass for neutron star is below 2${\rm M}_{\odot}$ for DM masses
$\gtrsim 10^{-8}$${\rm M}_{\odot}$.

{\it Acknowledgments} This research was supported by Ministerio de Ciencia e Innovaci\'on under
contracts FPA2013-43425-P as well as NewCompStar (COST Action MP1304).  LT
acknowledges support from the Ram\'on y Cajal Research Programme from
Ministerio de Ciencia e Innovaci\'on and from FP7-PEOPLE-2011-CIG under
Contract No. PCIG09-GA-2011-291679.


\begin{thebibliography}{9}


\bibitem{Betoule:2014frx} 
  M.~Betoule {\it et al.} [SDSS Collaboration],
  Astron.\ Astrophys.\  {\bf 568}, A22 (2014)

\bibitem{Ade:2013zuv} 
  P.~A.~R.~Ade {\it et al.}  [Planck Collaboration],
  Astron.\ Astrophys.\  {\bf 571}, A16 (2014)

\bibitem{ATLAS:2012ky} 
  G.~Aad {\it et al.} [ATLAS Collaboration],
  JHEP {\bf 1304}, 075 (2013)
  
\bibitem{Chatrchyan:2012me} 
  S.~Chatrchyan {\it et al.} [CMS Collaboration],
  JHEP {\bf 1209}, 094 (2012)
  
  
\bibitem{Klasen:2015uma} 
  M.~Klasen, M.~Pohl and G.~Sigl,
  arXiv:1507.03800 [hep-ph].
  
  
\bibitem{Ahmed:2010wy} 
  Z.~Ahmed {\it et al.}  [CDMS-II Collaboration],
  Phys.\ Rev.\ Lett.\  {\bf 106}, 131302 (2011)
  
\bibitem{Angloher:2011uu} 
 G.~Angloher {\it et al.},
  Eur.\ Phys.\ J.\ C {\bf 72}, 1971 (2012)
  
\bibitem{Aalseth} 
   C.~E.~Aalseth {\it et al.}  [CoGeNT Collaboration],
  Phys.\ Rev.\ Lett.\  {\bf 106}, 131301 (2011)
  
\bibitem{Bernabei:2010mq} 
  R.~Bernabei {\it et al.}  [DAMA and LIBRA Collaborations],
  Eur.\ Phys.\ J.\ C {\bf 67}, 39 (2010)

\bibitem{Akerib:2013tjd} 
  D.~S.~Akerib {\it et al.} [LUX Collaboration],
  Phys.\ Rev.\ Lett.\  {\bf 112}, 091303 (2014)

\bibitem{Felizardo:2011uw} 
  M.~Felizardo, T.~A.~Girard, T.~Morlat, A.~C.~Fernandes, A.~R.~Ramos, J.~G.~Marques, A.~Kling and J.~Puibasset {\it et al.},
  Phys.\ Rev.\ Lett.\  {\bf 108}, 201302 (2012)

\bibitem{Essig-Aprile} 
  R.~Essig, A.~Manalaysay, J.~Mardon, P.~Sorensen and T.~Volansky,
  Phys.\ Rev.\ Lett.\  {\bf 109}, 021301 (2012);
 E.~Aprile {\it et al.}  [XENON100 Collaboration],
  Phys.\ Rev.\ Lett.\  {\bf 109}, 181301 (2012)

\bibitem{Agnese:2013jaa} 
  R.~Agnese {\it et al.} [SuperCDMS Collaboration],
  Phys.\ Rev.\ Lett.\  {\bf 112},  241302 (2014)
  
  
\bibitem{Dai:2009ik} 
  D.~C.~Dai and D.~Stojkovic,
  JHEP {\bf 0908}, 052 (2009)
  
\bibitem{Kouvaris:2015rea} 
  C.~Kouvaris and N.~G.~Nielsen,
  arXiv:1507.00959 [hep-ph].



\bibitem{Tews:2012fj} 
  I.~Tews, T.~Krüger, K.~Hebeler and A.~Schwenk,
  Phys.\ Rev.\ Lett.\  {\bf 110},  032504 (2013)




\bibitem{Demorest:2010bx} 
  P.~Demorest, T.~Pennucci, S.~Ransom, M.~Roberts and J.~Hessels,
  Nature {\bf 467}, 1081 (2010)
  
\bibitem{Antoniadis:2013pzd} 
  J.~Antoniadis, P.~C.~C.~Freire, N.~Wex, T.~M.~Tauris, R.~S.~Lynch, M.~H.~van Kerkwijk, M.~Kramer and C.~Bassa {\it et al.},
  Science {\bf 340}, 6131 (2013)


\bibitem{Goldman:1989nd} 
  I.~Goldman and S.~Nussinov,
  Phys.\ Rev.\ D {\bf 40}, 3221 (1989)
  
\bibitem{Kouvaris:2011fi} 
  C.~Kouvaris and P.~Tinyakov,
  Phys.\ Rev.\ Lett.\  {\bf 107}, 091301 (2011)
  

\bibitem{Kouvaris:2010jy} 
  C.~Kouvaris and P.~Tinyakov,
  Phys.\ Rev.\ D {\bf 83}, 083512 (2011)


\bibitem{Kouvaris:2007ay} 
  C.~Kouvaris,
  Phys.\ Rev.\ D {\bf 77}, 023006 (2008)
  
  
\bibitem{Bertone:2007ae} 
  G.~Bertone and M.~Fairbairn,
  Phys.\ Rev.\ D {\bf 77}, 043515 (2008)
  
\bibitem{Kouvaris:2010vv} 
  C.~Kouvaris and P.~Tinyakov,
  Phys.\ Rev.\ D {\bf 82}, 063531 (2010)
  
\bibitem{McCullough:2010ai} 
  M.~McCullough and M.~Fairbairn,
  Phys.\ Rev.\ D {\bf 81}, 083520 (2010)


\bibitem{deLavallaz:2010wp} 
  A.~de Lavallaz and M.~Fairbairn,
  Phys.\ Rev.\ D {\bf 81}, 123521 (2010)
  
\bibitem{PerezGarcia:2011hh} 
  M.~A.~Perez-Garcia and J.~Silk,
  Phys.\ Lett.\ B {\bf 711}, 6 (2012)

\bibitem{PerezGarcia:2010ap} 
  M.~A.~Perez-Garcia, J.~Silk and J.~R.~Stone,
  Phys.\ Rev.\ Lett.\  {\bf 105}, 141101 (2010)
  



\bibitem{Li:2012ii} 
  A.~Li, F.~Huang and R.~X.~Xu,
  Astropart.\ Phys.\  {\bf 37}, 70 (2012)
  
\bibitem{Sandin:2008db} 
  F.~Sandin and P.~Ciarcelluti,
  Astropart.\ Phys.\  {\bf 32}, 278 (2009)
  
  
\bibitem{Leung:2011zz} 
  S.~C.~Leung, M.~C.~Chu and L.~M.~Lin,
  Phys.\ Rev.\ D {\bf 84}, 107301 (2011)
  
  
\bibitem{Xiang:2013xwa} 
  Q.~F.~Xiang, W.~Z.~Jiang, D.~R.~Zhang and R.~Y.~Yang,
  Phys.\ Rev.\ C {\bf 89},  025803 (2014)
  
\bibitem{Goldman:2013qla} 
  I.~Goldman, R.~N.~Mohapatra, S.~Nussinov, D.~Rosenbaum and V.~Teplitz,
  Phys.\ Lett.\ B {\bf 725}, 200 (2013)

\bibitem{Leung:2013pra} 
  S.-C.~Leung, M.-C.~Chu, L.-M.~Lin and K.-W.~Wong,
  Phys.\ Rev.\ D {\bf 87},  123506 (2013)


\bibitem{Spergel:1999mh} 
  D.~N.~Spergel and P.~J.~Steinhardt,
  Phys.\ Rev.\ Lett.\  {\bf 84}, 3760 (2000)

\bibitem{Harvey:2015hha} 
  D.~Harvey, R.~Massey, T.~Kitching, A.~Taylor and E.~Tittley,
  Science {\bf 347}, 1462 (2015)


\bibitem{Borucki:2010zz} 
  W.~J.~Borucki {\it et al.} [Kepler Collaboration],
  Science {\bf 327}, 977 (2010)




\bibitem{Alcock:1995zx} 
  C.~Alcock {\it et al.}  [MACHO Collaboration],
  Astrophys.\ J.\  {\bf 542}, 281 (2000)

\bibitem{Sumi:2011kj} 
   T.~Sumi, K.~Kamiya, A.~Udalski, D.~P.~Bennett, I.~A.~Bond, F.~Abe, C.~S.~Botzler and A.~Fukui {\it et al.},
  Nature {\bf 473}, 349 (2011)
  
\bibitem{Udalski:2004dy} 
  A.~Udalski,
  Acta Astron.\  {\bf 53}, 291 (2003)

\bibitem{Cassan:2012jx} 
   A.~Cassan {\it et al.},
  Nature {\bf 481}, 167 (2012)


\bibitem{Narain:2006kx} 
  G.~Narain, J.~Schaffner-Bielich and I.~N.~Mishustin,
  Phys.\ Rev.\ D {\bf 74}, 063003 (2006)

  
  
\bibitem{Shapiro:1983du} 
  S.~L.~Shapiro and S.~A.~Teukolsky,
  ``Black holes, white dwarfs, and neutron stars: The physics of compact objects,''
  New York, USA: Wiley (1983) 645 p
  
   
\bibitem{Kurkela:2014vha} 
  A.~Kurkela, E.~S.~Fraga, J.~Schaffner-Bielich and A.~Vuorinen,
  Astrophys.\ J.\  {\bf 789}, 127 (2014)

\bibitem{Negele:1971vb} 
  J.~W.~Negele and D.~Vautherin,
  Nucl.\ Phys.\ A {\bf 207}, 298 (1973)

\bibitem{Ruester:2005fm} 
  S.~B.~Ruester, M.~Hempel and J.~Schaffner-Bielich,
  Phys.\ Rev.\ C {\bf 73}, 035804 (2006)
  
  
 \bibitem{harrison-wheeler}
  B. K. Harrison, K. S. Thorne, M. Wakano and J. A. Wheeler, ``Gravitation Theory and Gravitational Collapse", The University of Chicago Press, 1965.
  
  
\bibitem{Nussinov:1985xr} 
  S.~Nussinov,
  Phys.\ Lett.\ B {\bf 165}, 55 (1985)

\bibitem{Kaplan:1991ah} 
  D.~B.~Kaplan,
  Phys.\ Rev.\ Lett.\  {\bf 68}, 741 (1992)
  
\bibitem{Hooper:2004dc} 
  D.~Hooper, J.~March-Russell and S.~M.~West,
  Phys.\ Lett.\ B {\bf 605}, 228 (2005)
  
\bibitem{Kribs:2009fy} 
  G.~D.~Kribs, T.~S.~Roy, J.~Terning and K.~M.~Zurek,
  Phys.\ Rev.\ D {\bf 81}, 095001 (2010)
  
  
  
  
  
  
\bibitem{Gorbunov:2011zzc} 
  D.~S.~Gorbunov and V.~A.~Rubakov,
  ``Introduction to the theory of the early universe: Cosmological perturbations and inflationary theory,''
  Hackensack, USA: World Scientific (2011) 489 p


\bibitem{Griest:2013aaa} 
  K.~Griest, A.~M.~Cieplak and M.~J.~Lehner,
  Phys.\ Rev.\ Lett.\  {\bf 111}, 181302 (2013);
  K.~Griest, A.~M.~Cieplak and M.~J.~Lehner,
  Astrophys.\ J.\  {\bf 786}, 158 (2014)


\bibitem{Tisserand:2006zx} 
  P.~Tisserand {\it et al.} [EROS-2 Collaboration],
  Astron.\ Astrophys.\  {\bf 469}, 387 (2007)

\bibitem{future}
L. Tolos and J. Schaffner-Bielich, {\it in preparation}
    
\end{thebibliography}
\end{document}


\title{Erratum: "Dark Compact Planets"}

\author{Yannick Dengler}
\email{dengler@itp.uni-frankfurt.de}
\affiliation{Institut f\"ur Theoretische Physik, J. W. Goethe Universit\"at, 
  Max von Laue-Str.~1, 60438 Frankfurt am Main, Germany}

\author{J\"urgen Schaffner-Bielich} 
\email{schaffner@astro.uni-frankfurt.de}
\affiliation{Institut f\"ur Theoretische Physik, J. W. Goethe Universit\"at, 
  Max von Laue-Str.~1, 60438 Frankfurt am Main, Germany}

\author{Laura Tolos} 
\email{tolos@ice.csic.es}
\affiliation{Institute of Space Sciences (ICE, CSIC), Campus UAB,  Carrer de Can Magrans, 08193 Barcelona, Spain; \\
Faculty  of  Science  and  Technology,  University  of  Stavanger,  4036  Stavanger,  Norway;\\
Institut d'Estudis Espacials de Catalunya (IEEC), 08034 Barcelona, Spain;\\
Frankfurt Institute for Advanced Studies, Ruth-Moufang-Str. 1, 60438 Frankfurt am Main, Germany}

\date{\today}

\maketitle

We have discovered an error in our code for solving the dimensionless Tolman-Oppenheimer-Volkov (TOV) equations for a system formed by ordinary matter (OM) and dark matter (DM) that only interact gravitationally. Following the work of Ref.~\cite{Narain:2006kx} for the determination of the dimensionless TOV equations for one specie,  the dimensionless TOV equations for two species read 
\begin{eqnarray}
&&\frac{dp'_{OM}}{dr}=-(p'_{OM}+\rho'_{OM})\frac{d \nu}{dr}, \nonumber \\
&&\frac{dm_{OM}}{dr}=4 \pi r^2 \rho'_{OM}, \nonumber \\
&&\frac{dp'_{DM}}{dr}=-(p'_{DM}+\rho'_{DM}) \frac{d \nu}{dr}, \nonumber \\
&&\frac{dm_{DM}}{dr}=4 \pi r^2 \rho'_{DM}, \nonumber \\
&&\frac{d \nu}{dr}=\frac{(m_{OM}+m_{DM}) + 4 \pi r^3(p'_{OM}+p'_{DM})}{r(r-2(m_{OM}+m_{DM}))}, 
\end{eqnarray}
where $p'$ and $\rho'$ are the dimensionless pressure and energy density of each specie. In order to obtain these dimensionless quantities for each specie one has to divide the physical quantities by one of the physical masses, that is, by the neutron mass or the dark matter particle
mass. We erroneously divided the physical quantities for ordinary matter by the neutron mass and those for dark matter by the dark matter particle mass. Given the freedom in choosing the particle mass, we have decided to use the dark matter particle mass, so that
 $p'_{OM}=P_{OM}/m_D^4$ and $\rho'_{OM}=\rho_{OM}/m_D^4$ as well as $p'_{DM}=P_{DM}/m_D^4$ and $\rho'_{DM}=\rho_{DM}/m_D^4$, with $m_D$ being the dark matter particle
mass.  The physical mass and radius for each specie are then given by $R_{OM}=
 (M_p/m_{DM}^2) \, r_{OM}$ and $M_{OM}= (M_p^3/m_{DM}^2) \, m_{OM}$  together with $R_{DM}=
 (M_p/m_{DM}^2) \, r_{DM}$ and $M_{DM}= (M_p^3/m_{DM}^2) \,  m_{DM}$, where $M_p$ is the Planck mass \cite{Narain:2006kx}.
 Therefore, the plots have changed, although the results and conclusions are qualitatively similar, as we discuss in the following. In particular, we still find dark compact planets with Earth-like or Jupiter-like masses, depending on the strength of the interacting dark matter, but with radii smaller by four orders of magnitude than those found in Ref.~\cite{Tolos:2015qra}

\begin{figure}[!t]
\includegraphics[width=0.5\textwidth,height=8cm]{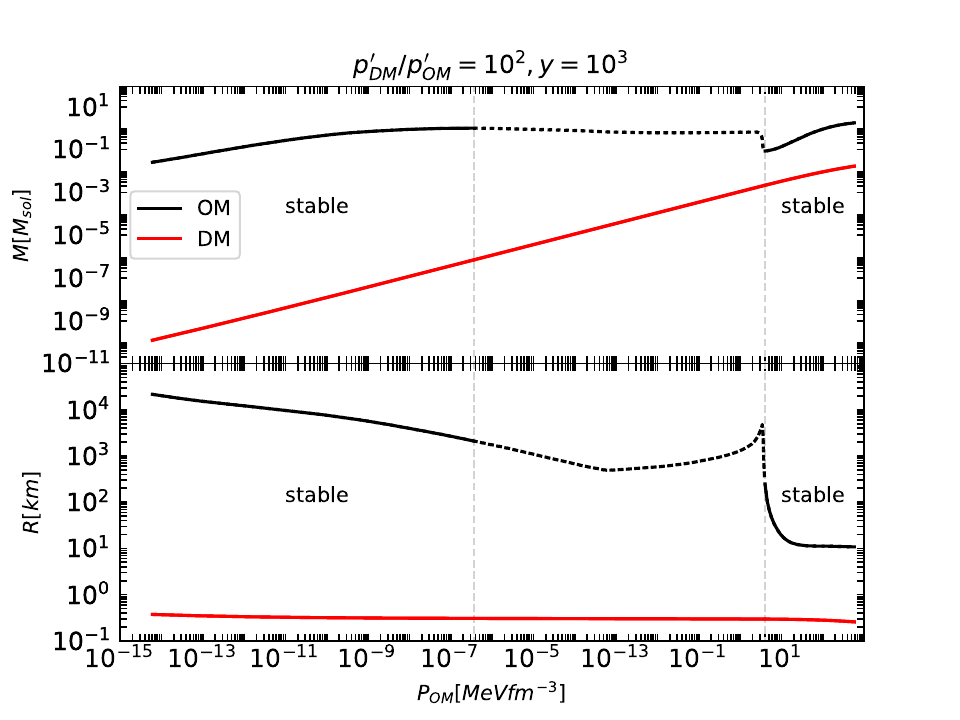}
\caption{Mass and radius of OM and DM objects as a function of the pressure of OM. The solid lines indicate
  the stable  regions for OM and DM, separately. The vertical
  dashed lines delimit the common stable regions for both species. The
  calculation is performed for the strongly interacting DM case, with a strength
  parameter $y=10^{3}$, and the ratio of dimensionless pressures between DM and OM is equal
  to $10^{2}$.}
\label{stability}
\end{figure}

\begin{figure}[!t]
\includegraphics[width=0.5\textwidth,height=8cm]{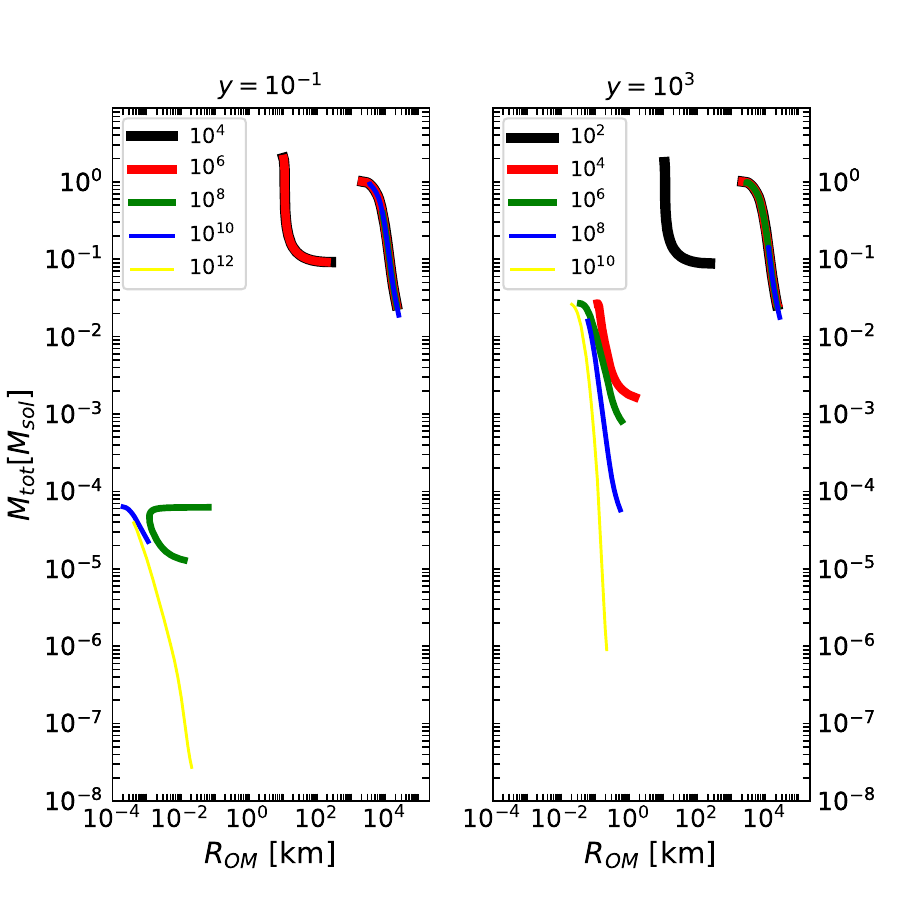}
\caption{The total mass of the dark compact object as a function of the
  observable radius (radius of OM) for different ratios of 
  pressures between DM and OM. Two different strengths
  for the interacting DM are used: $y=10^{-1}$ (left column) and $y=10^3$
  (right column).}
\label{rnm_mtot}
\end{figure}

\begin{figure}[!t]
\includegraphics[width=0.5\textwidth,height=7cm]{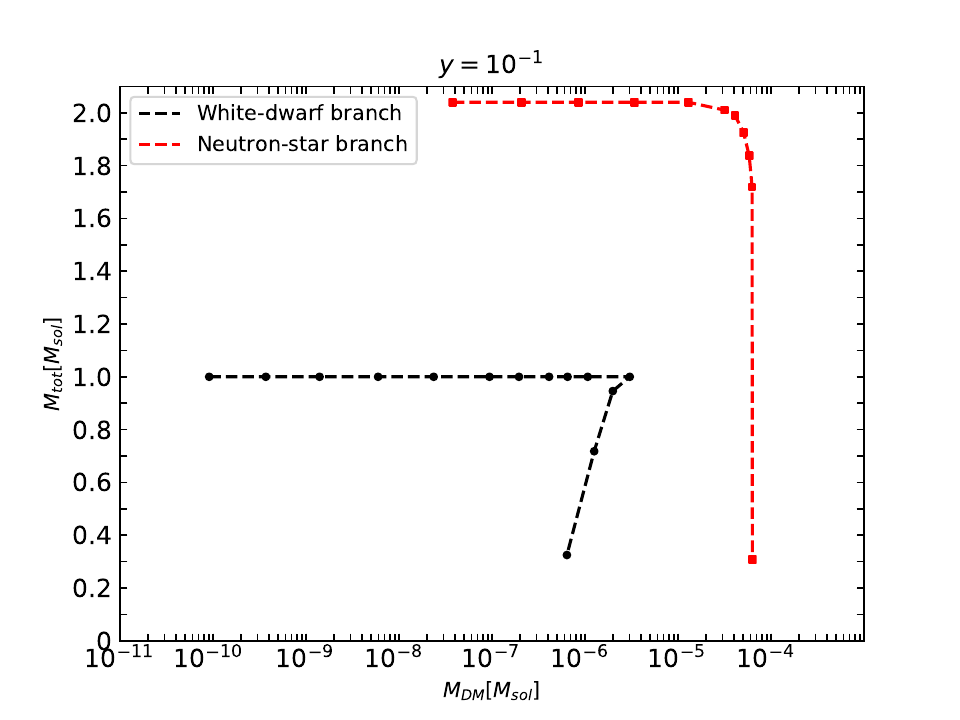}
\caption{The total mass of the dark compact object as a function of the mass
  of DM for weakly interacting DM, $y=10^{-1}$. The full squares correspond to the neutron-star branch, while the full circles indicate the white-dwarf branch.}
\label{mtot_mdar}
\end{figure}

In Fig.~\ref{stability} we present an example of the analysis of the stable configuration of both OM and DM. In this case, the
calculation is performed for the strongly interacting DM case, with a strength parameter $y=10^{3}$, and the ratio of dimensionless pressures between DM and OM is equal
  to $10^{2}$. The analysis of this figure is identical as the one performed when discussing Fig.~1 of Ref.~\cite{Tolos:2015qra}, although in the present case the dark matter component is always stable.

In Fig.~\ref{rnm_mtot} we present the corrected figure for the mass-radius relationships of the dark
compact objects for different values of the ratio of DM versus OM dimensionless
central pressures as well as the weakly and strongly interacting DM cases.  Our results are qualitatively the same as in Fig.~2 of Ref.~\cite{Tolos:2015qra},  although the stable mass-radius configurations appear at larger  $p'_{DM}/p'_{OM}$ ratios and the values of the observable radii (radius of ordinary matter) are four orders of magnitude smaller than those shown in that previous figure.  With the correct rescaling the gravitational effects
from DM on OM, in particular on white dwarf matter, those effects are now much more pronounced  leading to much smaller radii as compared to our previous calculations in Ref.~\cite{Tolos:2015qra}.
Two stable mass-radius configurations, that correspond to the neutron-star and white-dwarf branches
of OM, are observed for ratios below $p'_{DM}/p'_{OM}=10^8$ in the weakly interacting case ($y=10^{-1}$) and $p'_{DM}/p'_{OM}=10^4$ in the strongly interacting case ($y=10^{3}$).
As seen in Fig.~2 of Ref.~\cite{Tolos:2015qra} and corroborated here, only one branch survives for larger ratios of pressure, with densities
for OM below the neutron drip line for the largest pressure ratios, but unconventional masses and radii. Those are the dark compact planets (DCPs), with Earth-like masses of $M=$$10^{-4} -10^{-6}$${\rm M}_{\odot}$ or lower, but radii about one meter for the weakly interacting DM case. As for the strongly interacting case, these DCPs can reach masses similar
to the Jupiter mass of $10^{-2}-10^{-4}$${\rm M}_{\odot}$ or lower, and have radii of a kilometer. The DM content of these DCPs ranges from $M \sim $ $10^{-4}
-  10^{-8}$${\rm M}_{\odot}$ for Earth-like planets, and $M \sim $ $10^{-2}
-10^{-6}$${\rm M}_{\odot}$ for Jupiter-like ones.

Finally, in Fig.~\ref{mtot_mdar}, we display the corrected  total mass of the dark compact object as a function of
the mass of DM for the strength parameter $y=10^{-1}$ and for the
neutron-star and white-dwarf branches. This figure is similar to Fig.~3 of Ref.~\cite{Tolos:2015qra}, as the neutron star with 2 ${\rm M}_{\odot}$
can not be obtained as the DM
content reaches masses beyond $10^{-5}$${\rm M}_{\odot}$, one order of magnitude larger than the one obtained in Ref.~\cite{Tolos:2015qra}. A similar trend is obtained for the neutron-star branch in the strongly interacting DM case.
Comparing this new figure with Fig.~3 of Ref.~\cite{Tolos:2015qra}, in both cases there is still a more stringent limit in the DM content
coming from the reduction of the nominal mass of a white dwarf of 1 ${\rm M}_{\odot}$.  However, the white-dwarf branch of the present figure shows a reduction of the nominal mass with decreasing DM content. This is due to the fact that we can have two different OM density profiles with the same DM content. A similar behavior is found for the white-dwarf branch for the strongly interacting
DM case.